\documentstyle[pra,aps,amsfonts,multicol,graphicx]{revtex}

\newcommand{\be}[1]{\begin{equation} #1 \end{equation}}
\newcommand{\bea}[1]{\begin{eqnarray} #1 \end{eqnarray} }
\newcommand{\ba}[2]{\left(\begin{array}{#1}#2\end{array}\right)}

\newcommand{\C}{{\Bbb C}}
\newcommand{\tr}[1]{{\rm Tr}\left(#1\right)}

\newcommand{\qed}{\hfill$\Box$}

\newtheorem{theorem}{Theorem}

\draft
\title{A comparison of the entanglement measures negativity and concurrence}
\author{Frank Verstraete\cite{FV}, Koenraad Audenaert\cite{KA}, Jeroen Dehaene\cite{JD} and Bart De Moor\cite{BDM}}
\address{Katholieke Universiteit Leuven,
Department of Electrical Engineering, Research Group SISTA\\
Kard. Mercierlaan 94, B-3001 Leuven, Belgium }
\begin{document}

\pagestyle{plain} \pagenumbering{arabic}

\maketitle
\begin{abstract}
In this paper we investigate two different entanglement measures
in the case of mixed states of two qubits. We prove that the
negativity of a state can never exceed its concurrence and is
always larger then $\sqrt{(1-C)^2+C^2}-(1-C)$ where $C$ is the
concurrence of the state. Furthermore we derive an explicit
expression for the states for which the upper or lower bound is
satisfied. Finally we show that similar results hold if the
relative entropy of entanglement and the entanglement of
formation are compared.
\end{abstract}

\begin{multicols}{2}[]
\narrowtext

The concept of negativity originates from the observation due to
Peres \cite{peres} that taking a partial transpose of a density
matrix associated with a separable state is still a valid density
matrix and thus positive (semi)definite. Subsequently
M.Horodecki,P.Horodecki and R.Horodecki \cite{horodecki} proved
that this was a necessary and sufficient condition for a state to
be separable if the dimension of the Hilbert space does not
exceed $6$. In the case of an entangled mixed state two qubits,
the negativity is defined as two times the absolute values of the
negative eigenvalue of the partial transpose of a state.
Recently, Vidal and Werner proved that the negativity is an
entanglement monotone and therefore a good entanglement measure
\cite{VidalWerner}. Furthermore, the concept of negativity is of
importance as it leads to upper bounds for the entanglement of
distillation.

The concept of concurrence originates from the seminal work of
Hill and Wootters \cite{Hill,Wootters} where the exact expression
of the entanglement of formation of a system of two qubits was
derived. They showed that the entanglement of formation, an
entropic entanglement monotone, is a convex monotonic increasing
function of the concurrence.

Both measures have the same dimensionality and it is therefore a
natural question to compare them, as one is related to the concept
of entanglement of formation and the other one to the concept of
entanglement of distillation.

We will derive the possible range of values for the negativity if
the concurrence of the state is known. First of all we prove the
following conjecture by Eisert and Plenio\cite{Eisert}:
\begin{theorem} The negativity of an entangled mixed state of two
qubits can never exceed its concurrence.\end{theorem} To prove
this, we need the result of Wootters \cite{Wootters} that a state
with a given concurrence can always be decomposed as a convex sum
of four pure states all having the same concurrence. It is readily
checked that the negativity of a pure state is exactly equal to
its concurrence. Due to linearity of the partial trace operation,
the negativity of a mixed state is now obtained by calculating the
smallest eigenvalue of the matrix obtained by making the convex
sum of the partial transposes of the four pure states which have
all an equal negative eigenvalue. It is a well-known result due
to Weyl that the minimal eigenvalue of the sum of matrices always
exceeds the sum of the minimal eigenvalues, which concludes the
proof.\qed

The next step is to find the lowest possible value of the
negativity for given concurrence. To this end we need a
parameterization of the manifold of states with constant
concurrence. In \cite{Verstraetelorsvd}, it was shown how the
concurrence changes under the application of a LQCC operation of
the type
\begin{equation} \rho'=\frac{(A\otimes B)\rho(A\otimes
B)^\dagger}{{\rm Tr}\left((A\otimes B)\rho(A\otimes
B)^\dagger\right)}\end{equation}The transformation rule is:
\begin{equation} C(\rho')=C(\rho)\frac{|\det A||\det B|}{{\rm
Tr}\left((A\otimes B)\rho(A\otimes
B)^\dagger\right)}\end{equation}

It was furthermore shown that for each density matrix $\rho$ there
exists an $A$ and $B$ such that $\rho'$ is Bell diagonal. The
concurrence of a Bell diagonal state is only dependent on its
largest eigenvalue $\lambda_1$\cite{Hill}:
$C(\rho_{BD})=2\lambda_1(\rho_{BD})-1$. It is then straightforward
to obtain the parameterization of the surface of constant
concurrence (and hence constant entanglement of formation): it
consists of applying all complex full rank $2\times 2$ matrices
$A$ and $B$ on all Bell diagonal states with the given
concurrence, under the constraint that
\[{\rm Tr}\left(\left(\frac{A^\dagger
A}{|\det(A)|}\otimes\frac{B^\dagger B}{|\det
B|}\right)\rho\right)=1.\] It is clear that we can restrict
ourselves to matrices $A$ and $B$ having determinant 1 ($A,B\in
SL(2,C)$), as will be done in the sequel.

The extremal values of the negativity can now be obtained in two
steps: first find the state with extremal negativity for given
eigenvalues of the corresponding Bell diagonal state by varying A
and B , and then do an optimization over all Bell diagonal states
with equal $\lambda_1$.

The first step can be done by differentiating the following cost
function over the manifold of $A,B\in SL(2,C)$: \bea{
\Phi(A,B)&=&\lambda_{min}\left(\left((A\otimes
B)\rho_{BD}(A\otimes
B)^\dagger\right)^{\Gamma}\right)\\
&=&\lambda_{min}\left((A\otimes B^*)\rho_{BD}^{\Gamma}(A\otimes
B^*)^\dagger\right)} under the constraint
\[{\rm Tr}\left((A\otimes B^*)\rho_{BD}^\Gamma(A\otimes
B^*)^\dagger\right)=1,\] where the notation $\Gamma$ is used to
denote partial transposition.

There exists a very elegant formalism for differentiating the
eigenvalues of a matrix: given the eigenvalue decomposition of a
hermitian matrix $X=U\Lambda U^\dagger$, it is easy to proof that
$\dot{\Lambda}={\rm diag}(U^\dagger\dot{X}U)$, where 'diag' means
the diagonal elements of a matrix. We can readily apply this to
our Lagrange constrained problem. Indeed, the complete manifold of
interest is generated by varying  $A$ and $B$ as $\dot{A}=KA$ and
$\dot{B}=LB$ with $K,L$ arbitrary complex 2x2 traceless matrices
(the trace condition is necessary to keep the determinants
constant). Moreover the minimal eigenvalue is given by ${\rm
Tr}({\rm diag}[0;0;0;1]D)$ where $D$ is the diagonal matrix
containing the ordered eigenvalues of $C=PDP^\dagger=(A\otimes
B^*)\rho_{BD}^{\Gamma}(A\otimes B^*)^\dagger$ and $P$ the
eigenvectors of $C$. We proceed as
\bea{\small{\dot{\Phi}}&=&\small{\tr{P^\dagger
\dot{C}P\left(\underbrace{\ba{cccc}{0&0&0&0\\0&0&0&0\\0&0&0&0\\0&0&0&1}-\mu
I_4}_{=J(\mu)}\right)}}\nonumber\\
\small{\dot{C}}&=&\small{\left((K\otimes I_2)+(I_2\otimes
L)\right)C+C\left((K^\dagger\otimes I_2)+(I_2\otimes
L^\dagger)\right)}\nonumber} where $\mu$ is the Lagrange
multiplier. An extremum is obtained if $\dot{Phi}$ vanishes for
all possible traceless $K$ and $L$. Some straightforward algebra
shows that this condition is fulfilled iff
$CPJ(\mu)P^\dagger=P(DJ(\mu))P^\dagger$ is Bell diagonal (up to
local unitary transformations).

Next we have to distinguish two cases, namely when the Lagrange
multiplier $\mu=0$ and $\mu\neq 0$. The first case leads to the
condition that the eigenvector of $\rho^\Gamma$ corresponding to
the negative eigenvalue is a Bell state. It is indeed easily
checked that all density matrices with this property have
negativity equal to the concurrence, and this is clearly an
extremal case. We have therefore identified the class of states
for which the negativity is equal to the concurrence. It is
interesting to note that both all the pure states and all the Bell
diagonal states belong to this class.

The problem becomes much more subtle when the Lagrange multiplier
does not vanish. Using the arguments of the proof of theorem (5)
in \cite{Verstraetelorsvd}, it is easy to proof that the partial
transpose of an entangled state is always full rank and has at
most one negative eigenvalue: the set of equations (10-13) in
\cite{Verstraetelorsvd} is inconsistent with the constraints
$\lambda_3\leq 0$ and $\lambda_4<0$. $P(DJ(\mu))P^\dagger$ will
therefore be Bell diagonal either if the eigenvectors of $C$ are
Bell states, or possibly if  $DJ(\mu)$ contains eigenvalues with a
multiplicity of 2: in this last case the two eigenvectors
corresponding to the multiple eigenvalue are not uniquely  defined
and can be rotated to Bell states if the two other eigenvectors
were already Bell states. As the first case was already treated in
the previous paragraph, we concentrate on the second case.
Denoting the eigenvalues of $C$ as
$\lambda_1,\lambda_2,\lambda_3\geq 0\geq\lambda_4$, the
eigenvector corresponding to $\lambda_4$ can be different from a
Bell state iff we choose the Lagrange multiplier such that
$-\mu\lambda_3=(1-\mu)\lambda_4$. The eigenvectors corresponding
to $\lambda_1$ and $\lambda_2$ have to be Bell states. Therefore
all states for which the eigenvectors of the partial transposes
are , up to local unitary transformations,
of the form \be{P=\ba{cccc}{1/\sqrt{2}&1/\sqrt{2}& 0& 0\\
0& 0& 1& 0\\0&0&0&1\\1/\sqrt{2}&
-1/\sqrt{2}&0&0}\ba{cc}{I_2&0\\0&U_2}} with $U_2$ an arbitrary
2x2 unitary matrix will give extremal values of the negativity.
The next step is therefore to find the state belonging to this
class with minimal negativity for fixed concurrence, or
equivalently the one with the largest concurrence for fixed
negativity. Parameterizing the unitary $U$ as
$\ba{cc}{a&-b\\b^*&a^*}$, the class of states we are considering
is parameterized as:
\[\small{\ba{cccc}{\frac{\lambda_1+\lambda_2}{2}&0&0&ab(\lambda_3-\lambda_4)\\
0&\lambda_3|a|^2+\lambda_4|b|^2&\frac{\lambda_1-\lambda_2}{2}&0\\
0&\frac{\lambda_1-\lambda_2}{2}&\lambda_3|b|^2+\lambda_4|a|^2&0\\
a^*b^*(\lambda_3-\lambda_4)&0&0&\frac{\lambda_1+\lambda_2}{2}}}\]
The concurrence of this state can be calculated by finding the
Cholesky decomposition of $\rho=XX^\dagger$ and calculating the
singular values of $X^T(\sigma_y\otimes\sigma_y)X$. As $\rho$ is
a direct sum of two 2x2 matrices, this can be done exactly:
\bea{\sigma_1&=&\frac{\lambda_1+\lambda_2}{2}+|ab|(\lambda_3-\lambda_4)\\
\sigma_3&=&\frac{\lambda_1+\lambda_2}{2}-|ab|(\lambda_3-\lambda_4)\\
\sigma_2&=&\sqrt{(\lambda_3|a|^2+\lambda_4|b|^2)(\lambda_3|b|^2+\lambda_4|a|^2)}+\frac{\lambda_1-\lambda_2}{2}\\
\sigma_4&=&\sqrt{(\lambda_3|a|^2+\lambda_4|b|^2)(\lambda_3|b|^2+\lambda_4|a|^2)}-\frac{\lambda_1-\lambda_2}{2}}
The concurrence is therefore given by:
\be{\small{C=2(\lambda_3-\lambda_4)|ab|-2\sqrt{(\lambda_3|a|^2+\lambda_4|b|^2)(\lambda_3|b|^2+\lambda_4|a|^2)}}}
The task is now reduced to finding
$a,b,\lambda_1,\lambda_2,\lambda_3$ such that $C$ is maximized
for fixed $\lambda_4$. Some long but straightforward calculations
lead to the optimal
solution: \bea{|a|^2&=&1-|b|^2=\frac{\lambda_3}{|\lambda_4|}\\
\lambda_1&=&\lambda_2=\sqrt{\lambda_3|\lambda_4|}\\
1&=&\lambda_1+\lambda_2+\lambda_3+\lambda_4} This solution
corresponds to a state with two vanishing eigenvalues, while the
remaining two eigenvectors are a Bell state and a separable state
orthogonal to it:
\be{\rho=\ba{cccc}{C/2&0&0&C/2\\0&1-C&0&0\\0&0&0&0\\C/2&0&0&C/2}}
The concurrence $C$ is then related to the negativity
$N=2|\lambda_4|$ by the equation \be{N^2+2N(1-C)-C^2=0.} This
equation defines the lower bound we were looking for, as it
relates the minimal possible value of the negativity for given
concurrence. The state for which this minimum is reached is
special in the sense that it is a maximally entangled mixed state
\cite{ishizaka,Verstraetegenbell}: no global unitary
transformation can increase its entanglement. Moreover it is the
only mixed state that can be brought arbitrary close to a Bell
state by doing local operations (LOCC) on one copy of the state
only: it is a quasi-distillable state \cite{Verstraetegenbell}. We
have therefore proven:
\begin{theorem}
The negativity $N$ of a mixed state with given concurrence $C$ is
always smaller then $C$ with equality iff the eigenvector of
$\rho^\Gamma$  corresponding to its negative eigenvalue is a Bell
state (up to local unitary transformations). Moreover the
negativity is always larger then $\sqrt{(1-C)^2+C^2}-(1-C)$, with
equality iff the state is a rank 2 quasi-distillable state.
\end{theorem}

A scatter plot of the negativity versus the concurrence for all
entangled states is shown in figure 1.
\begin{figure}
\begin{center}
    \scalebox{0.4}{\includegraphics{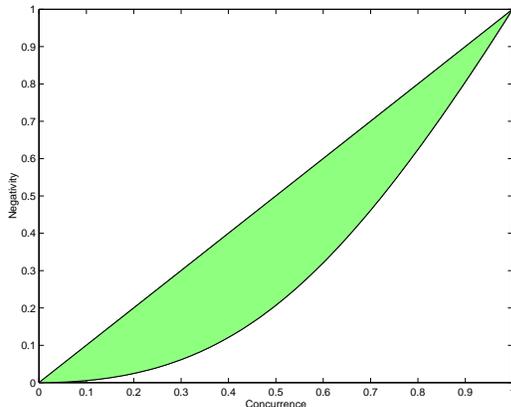}}\\
    \caption{Range of values of the negativity for given
    concurrence.}
\end{center}
\end{figure}

A similar analysis can be performed to compare the entanglement of
formation\cite{Wootters} and the relative entropy of
entanglement\cite{Vedral}. It is well-known that they coincide for
pure states, and that the relative entropy of entanglement can
never exceed the entanglement of formation. Due to the logarithmic
nature of these quantities however, finding the states with
minimal relative entropy of entanglement for given entanglement of
formation is very hard to do analytically. Numerical
investigations however showed that again the same
quasi-distillable rank 2 states minimize the relative entropy of
entanglement. It is indeed possible to show that these states are
local minima to the optimization problem. Using the results of
Verstraete et al.\cite{Verstraetegenbell}, this minimal value is
then given by: \be{E_R(\rho)=(C-2)\log(1-C/2)+(1-C)\log(1-C).} A
scatter plot of the range of values of the relative entropy of
entanglement is given in figure 2.
\begin{figure}
\begin{center}
    \scalebox{0.4}{\includegraphics{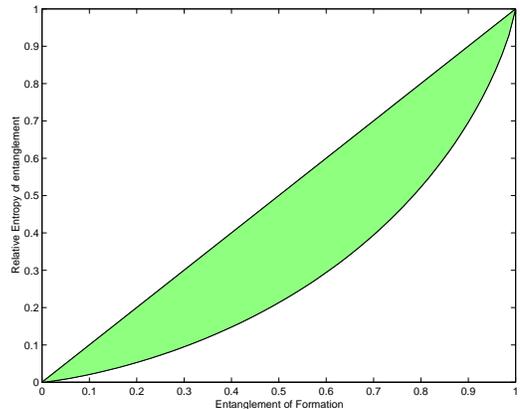}}\\
    \caption{Range of values of the Relative Entropy of Entanglement for given
    Entanglement of formation.}
\end{center}
\end{figure}
Both the relative entropy of entanglement and the negativity lead
to upper bounds on the entanglement of distillation. The strict
lower bounds for these quantities, derived in this paper, are
therefore nice illustrations of the expected irreversibility of
entanglement manipulations in mixed states.

\end{multicols}
\end{document}